\documentclass[10pt,aps,prl,reprint]{revtex4-1}
\usepackage[latin9]{inputenc}
\usepackage{amsmath}
\usepackage{amssymb}
\usepackage{graphicx}
\usepackage{esint}
\usepackage{bm,braket}

\makeatletter
\usepackage{color}

\makeatother

\begin{document}

\title{Nonlocal quasinormal modes for arbitrarily shaped three-dimensional plasmonic resonators}

\author{Mohsen Kamandar Dezfouli$^{1}$, Christos Tserkezis$^{2,3}$, N. Asger
Mortensen$^{2,3,4}$ and Stephen Hughes$^{1}$}

\affiliation{$^{1}$Department of Physics, Engineering Physics and Astronomy,
Queen's University, Kingston, Ontario, Canada K7L 3N6}

\affiliation{$^{2}$Department of Photonics Engineering, Technical University
of Denmark, {\O}rsteds Plads, 343, DK-2800 Kgs. Lyngby, Denmark}

\affiliation{$^{3}$Center for Nano Optics, University of Southern Denmark, Campusvej
55, DK-5230 Odense M, Denmark}

\affiliation{$^{4}$Danish Institute for Advanced Study, University of Southern Denmark, Campusvej 55, DK-5220 Odense M, Denmark}

\begin{abstract}
Nonlocal effects have been shown to be responsible for a variety of  non-trivial optical effects in small-size plasmonic nanoparticles, beyond classical electrodynamics.
However, it is not clear whether optical mode descriptions can be applied to such extreme confinement regimes. Here, we present a powerful and intuitive  quasinormal mode description of the nonlocal optical response for three-dimensional plasmonic nanoresonators. The nonlocal
hydrodynamical model and a generalized nonlocal optical response model for plasmonic nanoresonators are used to construct
an intuitive modal theory and to compare to the local Drude model
response theory. Using the example of a gold nanorod, we show how an efficient quasinormal mode picture
is able to accurately capture the blueshift of the resonances, the
higher damping rates in plasmonic nanoresonators, and the modified
spatial profile of the plasmon quasinormal modes, even at the single mode level. We exemplify the
use of this theory by calculating the Purcell factors of single quantum
emitters, the electron energy-loss spectroscopy spatial maps, as well as the Mollow triplet spectra of field-driven quantum dots with and without nonlocal effects for different size nanoresonators.
Our nonlocal quasinormal mode theory offers  a reliable and efficient technique to study both classical and quantum optical problems in nanoplasmonics.
\end{abstract}
\maketitle

\maketitle

\section{Introduction}
Fundamental studies of light-matter interactions using plasmonic devices
continue to make considerable progress and offering up a wide range of
applications \cite{Ciraci2012,Savage2012,Esteban2012,SEpeltonReview,Barbry2015,Hoang2016,AsgerOpinion}.
For spatial positions very close to metal resonators, the local Drude
model is known to fail, which challenges many of the usual modeling techniques
that use the classical Maxwell equations. In particular, charge density
oscillations become relevant, causing frequency shifts of the localized
surface plasmon (LSP) resonance as well as the appearance of additional
resonances above the plasmon frequency \cite{Ruppin,Fuchs,Leung,Abajo,McMahon2009,Raza2011}.
Such investigations have been performed using both density functional
theory (DFT) at the atomistic level \cite{Teperik2013}, and using
macroscopic nonlocal Maxwell's equations in the
form of the hydrodynamical model (HDM) \cite{Raza2011} and a generalized
nonlocal optical response (GNOR) model \cite{Mortensen2014}. However, so far,
with the exception of the simple cases of spherical or cylindrical nanoparticles,
nonlocal investigations have been primarily done using purely
numerical simulations \cite{Wiener2012,Stella2013,Filter2014}, which is not only computationally very extensive
for arbitrary shaped plasmonic systems, but also can lack important
physical insight; most of these calculations are also restricted to
2D geometries or simple particle shapes. Thus there is now need for more intuitive
and efficient formalisms with nonlocal effects included, for arbitrarily shaped metal resonators in a numerically feasible way.

In optics and nanophotonics, one of the most successful analytical approaches to most
resonator problems is to adopt a modal picture of the optical cavity
(e.g., in cavity-QED and coupled mode theory). Recently, it has been
also shown how quasinormal modes (QNMs) can quantitatively describe
the dissipative modes of both dielectric cavities and LSP resonances
\cite{Kristensen2014},  and even hybrid structures of metals and photonic
crystals \cite{KamandarDezfouli2017}. In contrast to ``normal modes,''
which are solutions to Maxwell's equations subjected to (usually)
fixed or periodic boundary conditions, QNMs are obtained with open
boundary conditions \cite{Leung1994}, and they are associated with
complex frequencies whose imaginary parts quantify the system losses.
These QNMs  require a more generalized normalization \cite{Leung1994,Kristensen2012,Sauvan2013,Kristensen2015,Bai2013,Muljarov2016},
allowing for accurate mode quantities to be obtained such as the effective
mode volume or Purcell factor \cite{purcell}, i.e., the enhanced spontaneous
emission (SE) factor of a dipole emitter. These QNMs are typically computed
numerically from the Helmholtz equation with open boundary conditions,
e.g., with perfectly matched layers (PMLs), whose solution can then
be used to construct the full photon Green function (GF) of the medium\textemdash a
function that is well known to connect to many useful quantities in
classical and quantum optics \cite{Ford1984195,Sipe:87,Agarwal2000,Wubs2004,Anger2006,steve2004,steve2007,VanVlack2012,Wei2013,Ge2014njp,Kamandaracs}.
The GF can also be used (and indeed is required) to compute electron
energy loss spectroscopy (EELS) maps for plasmonic nanostructures
\cite{Abajo2008,Rossouw2011,Nicoletti2011,Mohammadi2012,Boudarham2012,Husnik2013,Christensen2014,Raza2015,Horl2015,Ge2016,Hobbs2016,qeels2017},
which is a notoriously difficult problem in computational electrodynamics, especially for nanoparticles or arbitrary shape.
Despite these successes with QNMs, in the presence of nonlocal effects, it is
not known if such a mode description even applies.

In this work, we show that, somewhat surprisingly, QNMs can indeed
be obtained and used to construct the full system GF for complex 3D
plasmonic nanoresonators with nonlocal effects, and even a single
mode description is accurate over a wide range of frequencies and
spatial positions. We start by redefining the Helmholtz equation that
is usually solved to obtain the local QNMs \cite{Ge2014njp},
and then extend this approach to the
case of nonlocal systems using a generalized Helmholtz equation, which
is applicable to both HDM and GNOR models. A semi-analytical modal GF is then
used to perform Purcell factor calculations of dipole emitters positioned
nearby plasmonic gold nanorods (a structure for which there is no known analytical GF). We then show the
accuracy of the modal Purcell factors against fully vectorial dipole calculations,
also computed in the presence of the nonlocal corrections. The calculated
QNMs are also used to accurately quantify the effective mode volume
associated with coupling to quantum emitters,
that can be used, e.g., for quantifying single photon source figures
of merit \cite{Hoang2016,Aharonovich2016}. Additionally, we examine the size dependence of the nonlocal behavior by investigating nanorods of different sizes, verifying the anticipated LSP blueshifts \cite{Abajo} and damping with decreasing nanoparticle
size \cite{Mortensen2014}.  Next we use our QNM technique to efficiently
calculate the EELS maps for different sizes of nanoparticles \cite{Rossouw2011,Horl2015,Ge2016}.
Finally, to more rigorously show the benefit of our nonlocal modal picture for use in quantum theory of light-matter interaction, we study the behavior of the Mollow triplets of field-driven quantum dots (QDs) coupled to plasmonic resonators \cite{Ge2013}, under the influence of nonlocal effects.

\section{Cavity Mode approach to nonlocal plasmonics}
Without nonlocal corrections to the metal, the QNMs, $\tilde{\mathbf{f}}_{\mu}\left(\mathbf{r}\right)$,
can be defined as the solution to the Helmholtz equation with open
boundary conditions (such as PMLs), 
\begin{equation}
\boldsymbol{\nabla}\times\boldsymbol{\nabla}\times\tilde{\mathbf{f}}_{\mu}\left(\mathbf{r}\right)-\left(\dfrac{\tilde{\omega}_{\mu}}{c}\right)^{2}\varepsilon\left(\mathbf{r},\omega\right)\,\tilde{\mathbf{f}}_{\mu}\left(\mathbf{r}\right)=0,
\end{equation}
where $\varepsilon\left(\mathbf{r},\omega\right)$ is the relative
dielectric function of the system, and $\tilde{\omega}_{\mu}=\omega_{\mu}-i\gamma_{\mu}$
is the complex resonance frequency that can also be used to quantify
the QNM quality factor, $Q_{\mu}=\omega_{\mu}/2\gamma_{\mu}$. For
metallic regions, the dielectric function can be described using the
local Drude model, $\varepsilon_{{\rm MNP}}\left(\mathbf{r},\omega\right)=1-\omega_{p}^{2}/\omega\left(\omega+i\gamma_{p}\right),$
with $\hbar\omega_{p}=8.29\,{\rm eV}$ and $\hbar\gamma_{p}=0.09\,{\rm eV}$
for the plasmon frequency and collision rate of gold \cite{Ge2014},
respectively. However, when considering the nonlocal nature of the
plasmonic system, the electric field displacement relates to the electric
field through an integral equation rather than a simple proportionality
\cite{McMahon2009,McMahon2010}. In this \textit{nonlocal} case, a
modified set of equations \cite{Raza2011,Mortensen2014} can be used
to define nonlocal QNMs, $\tilde{\mathbf{f}}_{\mu}^{{\rm nl}}\left(\mathbf{r}\right)$,
 from 
\begin{align}
\boldsymbol{\nabla}\times\boldsymbol{\nabla}\times\tilde{\mathbf{f}}_{\mu}^{{\rm nl}}\left(\mathbf{r}\right)-\left(\dfrac{\tilde{\omega}_{\mu}^{{\rm nl}}}{c}\right)^{2}\tilde{\mathbf{f}}_{\mu}^{{\rm nl}}\left(\mathbf{r}\right) & =i\tilde{\omega}_{\mu}^{\rm nl}\mu_{0}\mathbf{J}_{\mu},\label{eq:nl1}\\
\xi^{2}\boldsymbol{\nabla}\left[\boldsymbol{\nabla}\cdot\mathbf{J}_{\mu}\right]+\tilde{\omega}_{\mu}^{{\rm nl}}\left(\tilde{\omega}_{\mu}^{{\rm nl}}+i\gamma_{p}\right)\mathbf{J}_{\mu} & =i\tilde{\omega}_{\mu}^{{\rm nl}}\omega_{p}^{2}\varepsilon_{0}\tilde{\mathbf{f}}_{\mu}^{{\rm nl}}\left(\mathbf{r}\right),\label{eq:nl2}
\end{align}
where $\mathbf{J}_{\mu}$ is the induced current density and $\xi$ is a
phenomenological length scale associated with the nonlocal corrections
\cite{Mortensen2014}. Indeed, $\xi^{2}=\beta^{2}+D\left(\gamma_{p}-i\omega\right)$,
where $\beta$ is the hydrodynamic parameter proportional to the electron
Fermi velocity, $v_{{\rm F}}=1.39\times10^{6}\,{\rm m/s}$, and $D=2.9\times10^{-4}\,{\rm m^{2}/s}$
\cite{Tserkezis2016a} is the diffusion constant associated with the
short-range nonlocal response. While $\xi$ in its full form represents
the GNOR model, we can simply switch to the HDM by neglecting the
diffusion.

Traditionally in cavity physics, the concept of effective mode volume,
$V_{{\rm eff}}$, plays a key role in characterizing the mode properties;
historically, $V_{{\rm eff}}$ quantifies the degree of light confinement
in optical cavities, and is normally defined at the modal antinode
where, e.g., a quantum emitter is typically placed. Even though for
plasmonic dimers one can reasonably choose the gap center as the place
to calculate the mode volume, for plasmonic resonators in general,
this simple picture of mode volume is ambiguous. However, one can
still quantify an effective modal volume, $V_{{\rm eff}}^{{\rm nl}}\left(\mathbf{r}\right)={\rm Re}\left\{ 1/n_{b}^{2}\left[\mathbf{\tilde{f}}_{\mu}^{{\rm nl}}\left(\mathbf{r}\right)\right]^{2}\right\} $
(same definition holds for the local QNM, only one uses $\mathbf{\tilde{f}}_{\mu}$)
\cite{Kristensen2014}, for rigorous use in Purcell's formula, which
is associated with coupling to emitters at different locations outside
(but typically near) the metal nanoparticle within a background medium
of refractive index $n_{b}$. Such a position-dependent mode volume
can then be used in a generalized Purcell factor, 
\begin{equation}
F_{{\rm P}}\left(\mathbf{r}\right)=\frac{3}{4\pi^{2}}\left(\frac{\lambda_{c}}{n_{b}}\right)^{3}\frac{Q}{V_{{\rm eff}}\left(\mathbf{r}\right)},\label{eq:purcell}
\end{equation}
to obtain the SE enhancement rate of a dipole emitter
placed at $\mathbf{r}$ around a cavity with the resonance wavelength
of $\lambda_{c}$ and quality factor of $Q$. The quantum emitter
is assumed to be on resonance and aligned in polarization with the
LSP mode.

Recent work has shown that QNMs, when obtained in  normalized form (as done in this work), accurately describe lossy plasmonic
resonators using the local Drude model \cite{Ge2014njp,KamandarDezfouli2017,Axelrod2017}.
Here, we extend such an approach to include the nonlocal effects by
considering the ansatz, 
\begin{equation}
\mathbf{G}^{{\rm nl}}_{\rm sc}\left(\mathbf{r}_{1},\mathbf{r}_{2};\omega\right)=\sum_{\mu}\dfrac{\omega^{2}}{2\tilde{\omega}_{\mu}^{{\rm nl}}\left(\tilde{\omega}_{\mu}^{{\rm nl}}-\omega\right)}\,\mathbf{\tilde{f}}_{\mu}^{{\rm nl}}\left(\mathbf{r}_{1}\right)\mathbf{\tilde{f}}_{\mu}^{{\rm nl}}\left(\mathbf{r}_{2}\right),\label{eq:GFexpansion}
\end{equation}
for the scattered GF, which is extremely useful as it can be immediately used
to obtain the full position and frequency dependence of the generalized Purcell factor (SE enhancement factor) for a dipole emitter polarized along ${\bf n}$:
\cite{Anger2006} 
\begin{equation}
F({\bf r};\omega)=1+\frac{6\pi c^{3}}{\omega^{3}n_{b}}\,\mathbf{n}\cdot{\rm Im}\{\mathbf{G}^{\rm nl}_{\rm sc}\left(\mathbf{r},\mathbf{r};\omega\right)\}\cdot\mathbf{n}.
\label{eq:pf}
\end{equation}
Note that in a single mode
regime, if at the peak of the resonance frequency, $\omega=2\pi c/\lambda_{c}$,
then \eqref{eq:pf} reduces to \eqref{eq:purcell}.

\section{Results and example applications}
In this section, a range of applications are presented to demonstrate the power and reliability of the QNM theory for optical and quantum optical investigations of plasmonic resonators, down to the nonlocal regime. We emphasize that, once the QNMs are calculated, as discussed in subsection \ref{subA}, all the example studies are performed in matters of seconds owing to the analytical power of the technique.

\subsection{Local versus nonlocal quasinormal modes}\label{subA}
To obtain the system QNMs for the nonlocal HDM/GNOR model defined
via \eqref{eq:nl1} and \eqref{eq:nl2}, we employ the frequency
domain technique discussed in Ref.~\cite{Bai2013} (used for the
local Drude model), where an inverse GF approach is used to return the QNM in normalized form without having to carry out any spatial integral. We extend this method by incorporating nonlocal corrections, both for the HDM method ~\cite{Toscano2012} as well as the more complete GNOR model \cite{Mortensen2014,Tserkezis2016}.
While the system GF using the notation of \cite{Bai2013} finds a different form, for consistency, we follow the ansatz of \eqref{eq:GFexpansion} to briefly explain the technique.

The basic idea is to use a dipole excitation at the location of interest, $\mathbf{r}_0$, having a dipole moment of $\mathbf{d}$, to numerically obtain the scattered Green function (as explained in more detailed later), $\mathbf{G}_{\rm sc}\left(\mathbf{r}_{0},\mathbf{r}_{0};\omega\right)$, and then reverse \eqref{eq:GFexpansion} to arrive at
\begin{equation}
\tilde{\mathbf{f}}_c\left(\mathbf{r}_{0}\right)\cdot\mathbf{d}=\sqrt{\frac{\mathbf{d}\cdot\mathbf{G}_{\rm sc}\left(\mathbf{r}_{0},\mathbf{r}_{0};\tilde{\omega}_c\right)\cdot\mathbf{d}}{A\left(\tilde{\omega}_c\right)}},\label{eq:fnut}
\end{equation}
where a single QNM, $\tilde{\mathbf{f}}_c\left(\mathbf{r}\right)$, with the complex frequency $\tilde{\omega}_c$ is considered, and we have defined $A(\omega)={\omega^{2}}/{2\tilde{\omega}_c\left(\tilde{\omega}_c-\omega\right)}$ for simplicity. The above quantity is in fact all one needs to perform an integration-free normalization for the QNM, and in practice it is calculated at frequencies very close to the QNM frequency (so is the QNM of \eqref{eq:QNM}) to avoid divergent behavior. When inserted back into \eqref{eq:GFexpansion}, one arrives at
\begin{equation}
\tilde{\mathbf{f}}_c\left(\mathbf{r}\right)=\frac{\mathbf{G}_{\rm sc}\left(\mathbf{r},\mathbf{r}_{0};\tilde{\omega}_c\right)\cdot\mathbf{d}}{\sqrt{A\left(\tilde{\omega}_c\right)\left[\mathbf{d}\cdot\mathbf{G}_{\rm sc}\left(\mathbf{r}_{0},\mathbf{r}_{0};\tilde{\omega}_c\right)\cdot\mathbf{d}\right]}},\label{eq:QNM}
\end{equation}
which provides the full spatial profile of the QNM, given that one also keeps track of the system response at all other locations, $\mathbf{G}_{\rm sc}\left(\mathbf{r},\mathbf{r}_{0};\omega\right)$, within the same simulation.

The numerical implementation is done using the frequency domain finite-element
solver COMSOL \cite{comsol}, where an electric current dipole source is used to excite the system and iteratively search for the QNM frequencies by monitoring the strength of the system response \cite{Bai2013}. To obtain the QNM, one obtains the scattered GF as the difference between two dipole simulation GFs at frequencies very close to the QNM frequency, with and without the metal nanoparticles \cite{Bai2013}, either in local or nonlocal case. The computed QNM can then provide the full spectral and spatial shape of the resonances involved. In our calculations,  a computational domain of $0.5\,\mu{\rm m}^{3}$
was used for all simulations with the maximum element size of $0.2\,{\rm nm}$
on the nanoparticle surface and $0.6\,{\rm nm}$ inside. The maximum
element size elsewhere is set to $33\,{\rm nm}$ to ensure convergent
results over a wide range of frequencies, and 10 layers of PML were
used. We have checked that these parameters provide accurate numerical
convergence for both local and nonlocal simulations done in this work.

\begin{figure}
\begin{center}
\includegraphics[width=.9\columnwidth]{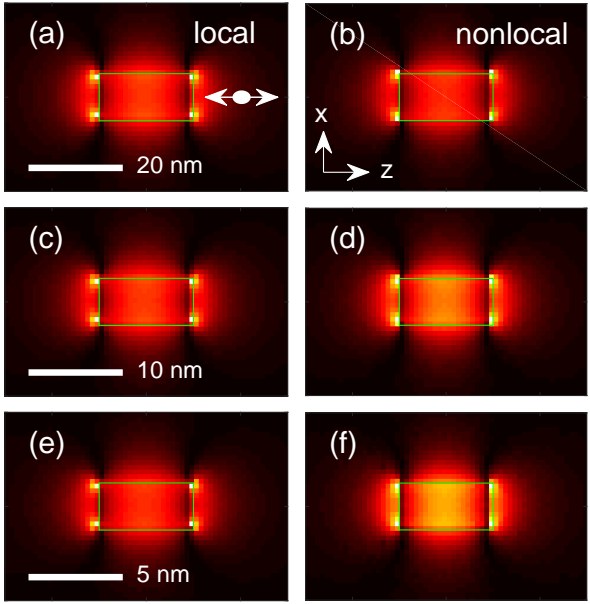}
\caption{\small{Comparison between the local QNM (a,c,e), $\left|\mathbf{\tilde f}(x,z,y_{0})\right|^{2}$,
and the nonlocal GNOR QNM (b,d,f), $\left|\mathbf{\tilde f}^{{\rm nl}}(x,z,y_{0})\right|^{2}$
for nanorods of different heights, $h=20\,{\rm nm}$, $h=10\,{\rm nm}$
and $h=5\,{\rm nm}$, where $y_{0}=0$ is at the center of the nanorods.
The same geometrical aspect ratio of 2 is used corresponding to a
radius of $r=5\,{\rm nm}$ for the largest resonator. The double arrow
in (a) shows the location of the dipole emitter at $10\,{\rm nm}$
away from the metallic surface, that is kept the same for all QNM calculations, and the green box represents the metallic border.
\label{fig:Comparison-qnm}}}
\end{center}
\end{figure}

Depicted in Fig.~\ref{fig:Comparison-qnm} are the computed QNMs
for three different gold cylindrical nanorods with the same aspect
ratio, varying from $20\,{\rm nm}$ to $5\,{\rm nm}$ in length (see
figure caption for details). The left panels represent the local Drude
model QNMs while the right panels show the QNMs using the nonlocal
GNOR model. As seen, the main QNM shapes are similar but a redistribution
of the localized field clearly occurs due to the inclusion of the
nonlocal corrections. While the local Drude model predicts a similar
mode shape for the different nanoparticle sizes, the nonlocal corrections
introduce a pronounced degree of modal reshaping for smaller nanoparticles.
Indeed, even for the largest nanoparticle shown in Figs.~\ref{fig:Comparison-qnm}(a,b),
higher field values are seen both inside as well as outside (but near)
the metallic region. 

\subsection{Purcell factors from coupled dipole emitters}

\begin{figure}[t]
\begin{center}
\includegraphics[width=0.88\columnwidth]{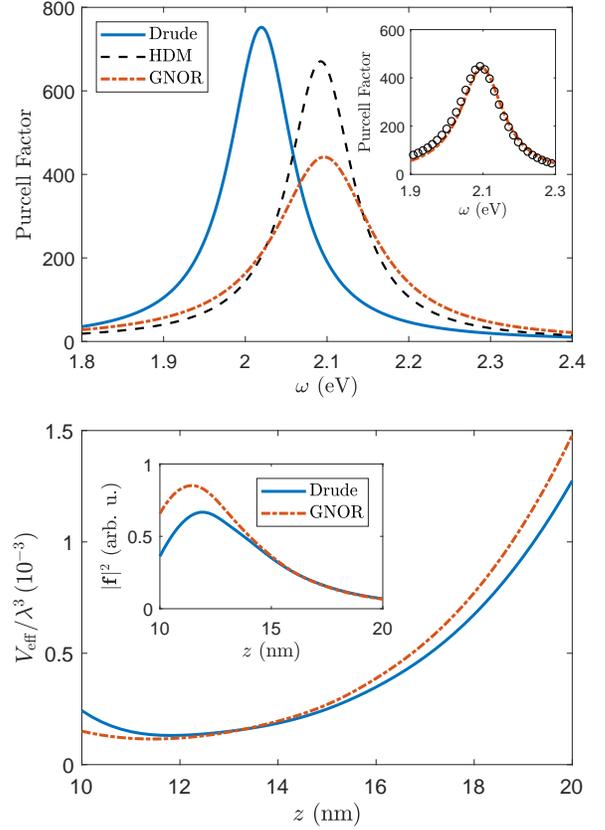}
\caption{\small{Top: Generalized Purcell factor for a dipole emitter placed $10\,{\rm nm}$ away form a gold nanorod of height $h=20\,{\rm nm}$ and radius of $r=5\,{\rm nm}$,
using Drude QNM, nonlocal HDM QNM and nonlocal GNOR QNM. The inset shows the agreement between full dipole calculations (with no
approximations) and GNOR QNM results. Bottom: The corresponding QNM effective mode volume, ${\rm V}_{{\rm eff}}$, is shown for a range of locations above the nanorod. Note that $z=10\,{\rm nm}$ is at the surface of the metallic nanoparticle. The inset shows
the modal absolute magnitude for completeness.
\label{fig:Comparison-pf}}}
\end{center}
\end{figure}

Figure~\ref{fig:Comparison-pf} shows the computed QNM Purcell factors
using the local Drude model as well as the two different nonlocal
models for the $h=20\,{\rm nm}$ nanorod. As can be seen, both HDM
and GNOR models predict the known blueshift of the plasmonic resonance
\cite{McMahon2009,Raza2011,Teperik2013}. However, the nonlocal prediction
of the peak enhancement strongly depends on the model chosen. The
GNOR model in particular predicts a considerably lower Purcell factor
due to the inclusion of diffusion, which accounts for surface-enhanced
Landau damping \cite{Mortensen2014}. Indeed, as will be discussed shortly, including the nonlocal effects modifies both
the quality factor and the mode volume associated with QNMs. The inset also shows the validity
of our Purcell factor calculations against full-dipole numerical calculations,
only shown for the nonlocal GNOR response. However, a similar degree
of agreement is observed for all other calculations both in Fig.~\ref{fig:Comparison-pf}
and what follows.

In Fig.~\ref{fig:Comparison-pf}, bottom panel, we additionally plot the corresponding
effective mode volume for a range of dipole locations, from the nanorod
surface (at $z=10$ nm) up to $10\,{\rm nm}$ away. A comparison
between the local Drude model (solid-blue) and the nonlocal GNOR (dashed-red)
is shown where a nontrivial difference is observed. Closer to the
metallic surface, smaller effective mode volumes are predicted by
the nonlocal corrections while further away the opposite takes place. The difference at larger distances however, is mainly due to nonlocal corrected resonant wavelengths that are used to normalize the mode volumes.

\begin{figure}
\begin{center}
\includegraphics[width=.9\columnwidth]{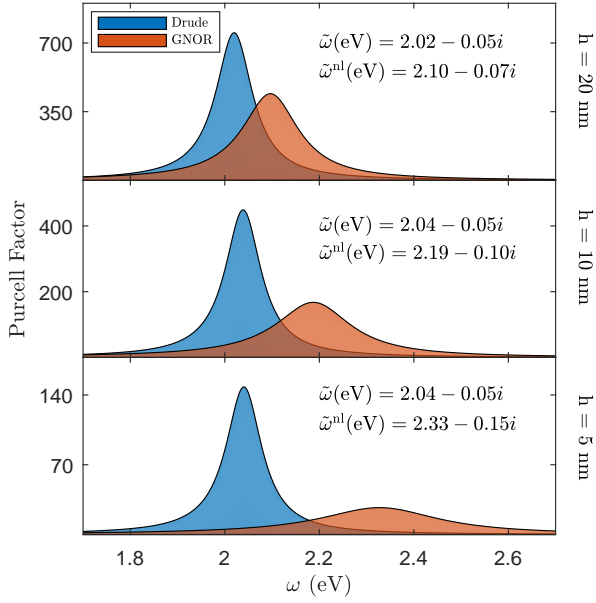}
\caption{\small{Size-dependent discrepancy in Purcell factor between the local Drude
model and the nonlocal GNOR model. Results are derived form analytical
QNM calculations when the dipole emitter is kept $10\,{\rm nm}$ away
along the z-axis. The complex resonance frequencies for both models are also shown in each case.\label{fig:Comparison-size}}}
\end{center}
\end{figure}

We also consider enhanced SE from the three
different gold nanoparticles discussed in Fig.~\ref{fig:Comparison-qnm}.
Plotted in Fig.~\ref{fig:Comparison-size}
are the QNM calculations of the Purcell factors for dipole emitters
located $10\,{\rm nm}$ away from nanorods, on the $z$-axis.
In each case, the local Drude results are compared with the nonlocal
GNOR results. Clearly, nonlocal corrections result in both
larger resonance blueshifts and larger damping rates (lower quality factors).

\subsection{Computing  EELS spatial maps}

\begin{figure}
\begin{center}
\includegraphics[width=.9\columnwidth]{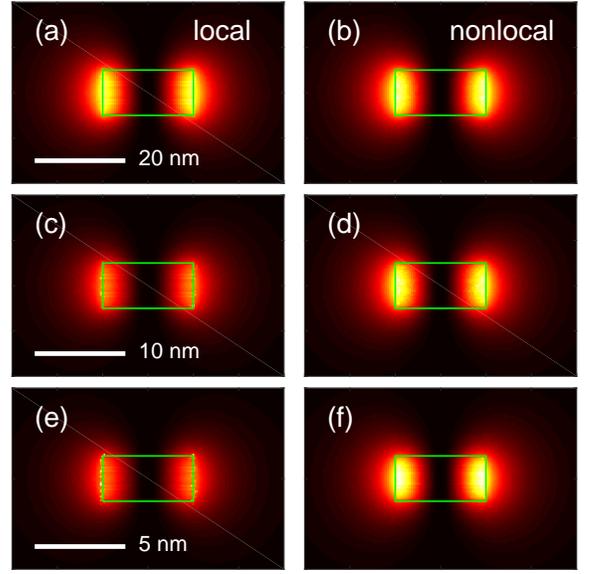}
\caption{\small{Comparison between the EELS map of the plasmonic nanorod using the
local Drude model, (a,c,e), and the nonlocal GNOR model, (b,d,f), where each map is calculated at the corresponding plasmonic peak frequency.
Same geometries as in Fig.~\ref{fig:Comparison-qnm} are used and
the green box represents the metallic border.\label{fig:Comparison-eels}}}
\end{center}
\end{figure}

For our next application of the nonlocal QNM theory,
we calculate the spatial maps associated with EELS experiments that
are obtained by nanometer-scale resolution in microscopy of LSP resonances
\cite{Rossouw2011,Nicoletti2011,Husnik2013,Hobbs2016,qeels2017}.
Since the GF is available at all locations through QNM expansion of
\eqref{eq:GFexpansion}, the EELS spectra in the $xz$ plane
subjected to an electron beam propagating along the $y$ axis can be easily
obtained from \cite{Rossouw2011,Ge2016}, 
\begin{align}
 & \Gamma\left(x,z;\omega\right)=\\
 & \quad-\frac{4e^{2}v^{2}}{\hbar}\int dtdt^{\prime}{\rm Im}\left\{ e^{i\omega\left(t^{\prime}-t\right)}G_{yy}\left(\mathbf{r}_{e}\left(t\right),\mathbf{r}_{e}\left(t^{\prime}\right);\omega\right)\right\} ,\nonumber 
\end{align}
where $v$ is the speed of electrons, and the single mode expansion for our
Green function---that is already confirmed to be very accurate---is used. The EELS calculations for all
three nanoparticles (of Fig.~\ref{fig:Comparison-qnm}) are shown
in Fig.~\ref{fig:Comparison-eels},  all computed at the corresponding plasmonic peaks frequencies. Note that there are some noticeable
numerical problems around the sharp corners of the metallic nanorod
when using the conventional Drude model theory on the left (which
is a known problem \cite{sharp2,sharp1}). Using the same meshing
scheme, however, the nonlocal description evidently helps to avoid
such nonphysical predictions. More importantly, as the nanoparticle
size decreases the EELS map becomes brighter at its maximum location
which originates from the higher modal amplitudes of the QNMs discussed
in Fig.~\ref{fig:Comparison-qnm}. We stress that with the computed
QNMs, such EELS maps are calculated instantaneously, which is a far
cry from the usual brute force numerical solvers. 

\subsection{Field-driven Mollow triplets and quantum optics regime}

Finally, in addition to the previous discussions on Purcell factor and mode volume that are essential for building quantum optical models of light-matter interaction \cite{Bozhevolnyi2017}, we discuss the quantum regime of field-driven Mollow triplets for QDs coupled to plasmonic nanoparticles. 
In the dipole and rotating wave approximations,
the total Hamiltonian of the coupled system can be written as \cite{Welsch1999,Ge2013}
\begin{align}
&H = \hbar \int d{\bf r} \int_0^\infty d\omega  \, \omega \,{\bf \hat f}^\dagger({\bf r},\omega) {\bf \hat f}({\bf r},\omega) +\hbar\omega_{x} \sigma^+ \sigma^- \\
&-\left [\sigma^+ \int_0^\infty \! d\omega\, {\bf d} \cdot {\bf \hat E}({\bf r}_{d},\omega) + {\rm H.c.}\right ] +
\frac{\hbar\Omega}{2}\left(\sigma^+e^{-i\omega_L}+\sigma^-e^{i\omega_L}\right) \nonumber,
\end{align}
where $\Omega=\braket{{\bf \hat E}_{\rm pump}({\bf r}_d)}\cdot{\bf d}/\hbar$ is the effective Rabi field, $\sigma^+,\sigma^-$ are the Pauli operators of the two-level atom (or exciton), $\omega_x$ is the resonance of the exciton,
${\bf d}$ is the dipole of the exciton, and  ${\bf \hat f},{\bf \hat f}^\dagger$ are the boson field operators.
Following the approach in
Ref.~\cite{Ge2013}, and using the interaction picture at the laser frequency $\omega_L$, one can derive a self-consistent generalized master equation in the 2nd-order Born-Markov approximation:
\begin{align}
\label{eq:ME1}
  \frac{\partial {\rho}}{\partial t}    & =
  \frac{1}{i\hbar}[H_S,\rho] + \int_0^t d\tau   \left \{{\tilde J_{\rm ph}(\tau)}   [-\sigma^+ \sigma^-(-\tau)\rho\, +  \right . \nonumber \\
& \left .  \phantom{\tilde J_{\rm ph}} +\sigma^-(-\tau)\rho\sigma^+  ] + {\rm H.c.} \right \},
\end{align}
where
 $\tilde J_{\rm ph}(\tau) =  \int_0^\infty d\omega J_{\rm ph}(\omega)   e^{i(\omega_L-\omega)\tau}$, with
 the photon-reservoir spectral function given by
$J_{\rm ph}(\omega) \equiv \frac{{\bf d} \cdot {\rm Im}[{\bf G}({\bf r}_{ d},{\bf r}_{ d};\omega)] {\bf  \cdot d}}{\pi\hbar\varepsilon_0}$, and
the time-dependent operators are defined through
 $\sigma^{\pm}(-\tau)=e^{-iH_S\tau/\hbar} \sigma^{\pm} e^{iH_S\tau/\hbar}$,
 with $H_S =  \hbar(\omega_{x}-\omega_L)\sigma^+\sigma^-
+ \hbar\Omega/2(\sigma^++\sigma^-)$,
 which results in a complex interplay between the values of the local density of states at the field-driven dressed states.
 Solving the master equation, and exploiting the quantum regression theorem, one can compute the incoherent spectrum of the QD emission from \cite{Ge2013}
\begin{align}
 S_0\left(\omega\right) & =  \lim_{t\rightarrow\infty} \textrm{Re}\left[\int_0^{\infty} d\tau \left(\left\langle\sigma^+(t-\tau)\sigma^-(t)\right\rangle\right.\right.\\
 & - \left.\left.\left\langle\sigma^+(t)\rangle \langle \sigma^-(t)\right\rangle\right)e^{i(\omega_L-\omega)\tau}\right],\nonumber
\end{align}
as well as the detected spectrum, which includes quenching and propagation from QD at $\mathbf{r}_0$ to a point detector at $\mathbf{r}_D$, from \cite{Ge2013}
\begin{align}
 S\left(\mathbf{r}_D;\omega\right) = \frac{2}{\varepsilon_0}\,|\mathbf{G}(\mathbf{r}_D,\mathbf{r}_0;\omega)\cdot\mathbf{d}|^2 \, S_0(\omega).
\end{align}
\begin{figure}
\begin{center}
\includegraphics[width=.9\columnwidth]{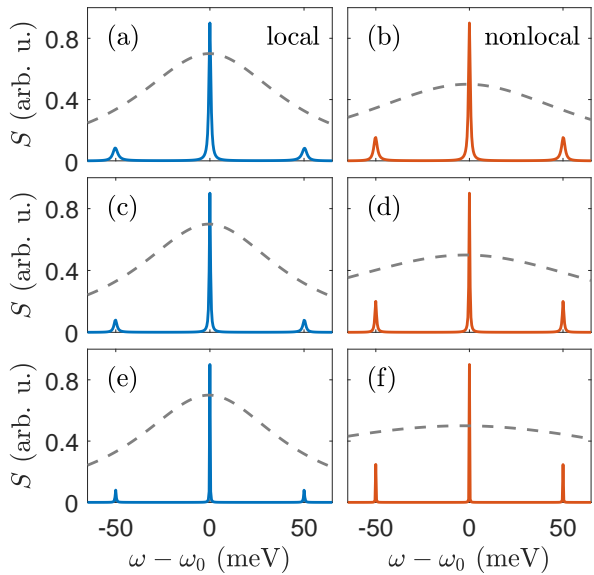}
\caption{\small{Detected spectra ($S$ evaluated at ${\bf r}_D=200\,\textrm{nm}$) of a field-driven QD coupled to plasmonic nanoparticles, where the same ordering of the particle size and QD location as in Fig.\,\ref{fig:Comparison-qnm} is followed, and we use an effective Rabi field of $\Omega=50\,\mbox{meV}$. The plasmonic enhancement is also shown in dashed-gray in background. Nonlocal investigations on the right predict relatively stronger side peaks for the Mollow triplet with narrower linewidths.\label{fig:Comparison-mollow}}}
\end{center}
\end{figure}

For example calculations, we assume a QD with the dipole moment of $|\mathbf{d}|=50\,\mbox{Debye}$ at $10\,\textrm{nm}$ away from the nanoparticle surface, at $x$=$0$. In particular, as with the  calculations above, we compare the local Drude model versus nonlocal GNOR model as shown in Fig.\,\ref{fig:Comparison-mollow}. As can be recognized, including the nonlocal effects, in general, predicts narrower linewidths for the Mollow  triplets (see e.g. Table\,\ref{tab:linewidth}) where the relative strength of the side peaks are also increased. This is attributed to the modified plasmonic enhancment in the nonlocal description as confirmed in Fig.\,\ref{fig:Comparison-size}, which can be also rigorously confirmed using analytical equations for the linewdiths derived in \cite{Ge2013}. It should be noted that, in general, the detected spectrum $S$ for the Mollow triplet problem can be different than the emitted $S_0$ as discussed in \cite{Ge2013}, however in our particular case, under the resonant excitation, we find that they have the same qualitative shape (and differ only quantitatively). We also stress that, these spectral calculations, at any detector position, can be trivially performed using a standard desktop through use of semi-analytical GF of \eqref{eq:GFexpansion}, demonstrating the power of our approach for carrying out complex problems in quantum optics.

\begin{table}
\begin{centering}
\begin{tabular}{|c|c|c|}
\hline 
  $h$ (nm) & Drude FWHM (meV) & GNOR FWHM (meV) \tabularnewline
\hline 
\hline 
20 & 1.28  & 1.21 \tabularnewline
\hline
10 & 0.74  & 0.61 \tabularnewline
\hline
5 & 0.45  & 0.42 \tabularnewline
\hline
\end{tabular}
\par\end{centering}
\caption{Linewidth of the central Mollow peak for the three nanoparticles, using both Drude model and GNOR model. Same trend holds for the side peaks.}
\label{tab:linewidth}
\end{table}

\section{Conclusions}
 We have presented an efficient and accurate modal description
of the nonlocal response of arbitrarily shaped metallic nanoparticles,
using a fully 3D model. We have shown how analytical nonlocal QNMs can be used
to accurately construct the system GF from which modal quantities
of interest such as Purcell factor and effective mode volume can be
derived. As anticipated, we first observe  the blueshift as well as the larger damping
rate for the LSP as a consequence of nonlocal effects. We further
confirmed the validity of our approach for different nanoparticle
sizes with full dipole solutions of the modified Maxwell equations, and we were able to successfully predict the size-dependent
nonlocal modifications with ease. As example applications of the theory, we described how our nonlocal QNMs
can be used to efficiently model Purcell factors of dipole emitters, EELS spatial maps as well as  Mollow triplet spectra of field-driven QD. The presented model has many
applications in both classical and quantum nanoplasmonics, offers considerable analytical insight into complex nonlocal problems, and could help pave
the way for a quantum description of both light and matter.

\section*{Funding Information}
We acknowledge funding from Queen's University, the Natural Sciences
and Engineering Research Council of Canada, the People Programme (Marie
Curie Actions) of the European Union's Seventh Framework Programme
(FP7/2007-2013) under REA grant agreement number 609405 (COFUNDPostdocDTU),
Villum Fonden (VILLUM Investigator grant no. 16498) and the Danish National Research Council (DNRF103).

\end{document}